\documentclass[a4paper]{jpconf}
\usepackage{graphicx}

\usepackage{graphics}
\usepackage{graphicx}
\usepackage{epsfig}
\usepackage{pstricks}
\usepackage{color}
\newcommand \beq{\begin{eqnarray}}
\newcommand \eeq{\end{eqnarray}}
\newcommand \bea{\begin{eqnarray}}
\newcommand \eea{\end{eqnarray}}
\newcommand \kvec{{\bf k}}
\newcommand \qvec{{\bf q}}

\def\simge{\mathrel{%
       \rlap{\raise 0.511ex \hbox{$>$}}{\lower 0.511ex \hbox{$\sim$}}}}
\def\simle{\mathrel{
       \rlap{\raise 0.511ex \hbox{$<$}}{\lower 0.511ex \hbox{$\sim$}}}}

\begin{document}

    \title{Finite-size analysis of the Fermi liquid properties of the homogeneous electron gas}

\author{Markus Holzmann$^{a,b*}$, Bernard Bernu$^a$, and David M. Ceperley$^c$}
\address{$^a$LPTMC, UMR 7600 of CNRS, Universit\'e Pierre et Marie Curie, 4 Place Jussieu,
75005 Paris, France\\
$^b$Univ. Grenoble 1/CNRS, LPMMC UMR 5493, Maison des Magist\`{e}res, 38042 Grenoble, France\\
$^c$Dept. of Physics and NCSA, U. of Illinois at
Urbana-Champaign, Urbana, IL 61801, USA}
\ead{$^*$markus@lptmc.jussieu.fr}

\begin{abstract}
We analyze the extrapolation to the thermodynamic limit of
Fermi liquid properties of the homogeneous electron gas in two
and three dimensions. Using field theory, we explicitly
calculate finite-size effects of the total energy, the
renormalization factor, and the effective mass at the Fermi
surface within the random phase approximation (RPA) and discuss
the validity for general metallic systems.

\end{abstract}

\section{Introduction}
We consider the Fermi liquid phase of the (unpolarized)
homogeneous electron gas in two and three dimensions ($d=2,3$)
at zero temperature, $T=0$. Quantum Monte Carlo (QMC)
calculations \cite{QMC3D,QMC2D} have not only provided the most
precise evaluations of the ground state energy, $E$, as a
function of density, $n$, but also give access to spectral
properties, \emph{e.g.} the momentum distribution, $n_k$, the
effective mass, $m^*$, of quasiparticle excitations, and the
corresponding spectral weight, the renormalization factor $Z$
\cite{QMC2D,DrummondNk,Momk2D,Ortiz,Na,Momk3D,Kwon1,Kwon2,Drummondmstar}.
Whereas QMC provides a powerful strategy to quantify
correlations in Fermi systems, it still suffers from two
fundamental limitations: the fixed-node error, and the
extrapolation to the thermodynamic limit. As shown recently,
calculations of the Fermi liquid parameters are drastically
affected by size effects \cite{Momk2D,Momk3D}, since typical
QMC computations are done with $N \simle 10^3$ electrons, but
extrapolation schemes based on analytical knowledge of
underlying properties significantly reduce the finite size
error \cite{FSE,DrummondFSE,Momk2D,Momk3D}. Here, to provide
further insight, we address the issue of extrapolating
calculations of finite (small) systems to the infinite volume
limit within common language of field theory. Strategies of
improving the many-body wave function to reduce the fixed-node
error in QMC are discussed elsewhere \cite{release,He,Hesaverio}.

At high densities, $r_s \to 0$,  the random phase
approximations (RPA) is expected to be valid, and we explicitly
calculate leading order size corrections within this
approximation, and discuss their general validity. Here,
$r_s=a/a_B$ with  $a= [2 (d-1)\pi n/d]^{-1/d}$ is the mean
inter particle distance, $a_B=\hbar^2/me^2$ the Bohr radius,
and $e$ and $m$ are the charge and mass of the electrons. Our
results agree with the leading order corrections in
\cite{FSE,Momk2D,Na,Momk3D} where finite size corrections have
been derived from a formulation directly based on the
many-body wave function. In the following, we explicitly
concentrate on three fundamental properties, the total energy,
$E$, the renormalization factor, $Z=n_{k_F^-}-n_{k_F^+}$, and
the effective mass, $m^*$, at the Fermi surface, $k_F$.

\section{Formalism}

From the dielectric function, $\epsilon(q,\omega)$, the ground
state energy per particle, $E$, inside a box of volume $V$ with
periodic boundary conditions, and $n=N/V$, can be obtained via
coupling constant integration \cite{NozieresPines} \beq
E=E_0+\frac{1}{2V} \sum_{\qvec \ne 0} \left(-v_q
-\frac{\hbar}{n} \int_0^1 d \lambda  \int_0^\infty
\frac{d\omega}{\pi} \left[
\frac{1}{\epsilon(q,i\omega;\lambda)}-1 \right] \right)
\label{Etot} \eeq where $v_q=2(d-1)\pi e^2/q^{d-1}$ is the
Coulomb potential and \beq E_0=\frac{2}{N} \sum_{\kvec \le k_F}
\varepsilon_k \label{Ekin} \eeq is the ideal gas kinetic energy
with $\varepsilon_k=\hbar^2 k^2/2m$. In the following we set
$\hbar=1$. Within the RPA \cite{BohmPines,Nozieresbook}, the
dielectric function is given in terms of the ideal-gas
density-density response function, $\chi_0(k,\omega)$, \beq
\epsilon_{RPA}(k,\omega)=1-v_k \chi_0(k,\omega) \eeq and the
coupling constant integration can be done explicitly \beq
E_{RPA}= E_0 +\frac{1}{ N} \sum_{\qvec \ne 0} \left[
-\frac{v_q}{2n}+ \int_0^\infty \frac{d\omega}{2\pi} \log
\epsilon_{RPA}(\qvec,i\omega) \right] \label{ERPA} \eeq

Spectral properties are obtained from the single particle
Green's function, $G(k,\omega)$, which equals
 \beq
G^{-1}(k,\omega)= \omega+ \mu -\varepsilon_k
-\Sigma(k,\omega)\eeq
 where the self energy, $\Sigma(k,\omega)$
accounts for interaction effects, and $\mu$ is the chemical
potential which fixes the Fermi energy, $\varepsilon_F=\mu$,
via
$\varepsilon_F=\varepsilon_{k_F}+\Sigma(k_F,0;\mu\equiv\varepsilon_F)$.
From  a frequency integration over the Green's function, the
momentum distribution, $n_k$, can be obtained, and the
magnitude of the jump at the Fermi surface,
$Z=n_{k_F^-}-n_{k_F^+}$, is directly related to the residual of
$G$ at ($k_F$,$\varepsilon_F$). In general, poles of the
Green's function give rise to quasi-particle excitations, whose
energies are given in terms of an effective mass $m^*$ at the
Fermi surface.

Within the RPA, the self energy equals
\cite{Rice,Quinn,Schulze} \beq \Sigma_{RPA}(k,\omega) &=&
 - \frac{1}{V}  \sum_{\qvec \ne 0}
 \left\{
 \frac{v_q}{2}
 \nonumber
 +
 \frac{v_q}{\epsilon(q,\varepsilon_{\kvec+\qvec}-\omega)}
 \left[
 \theta(\varepsilon_F-\varepsilon_{\kvec+\qvec})
 -\theta(\omega-\varepsilon_{\kvec+\qvec})
 \right]
 \right.
 \nonumber
 \\
 &&
 \left. +
 \int_{-\infty}^\infty \frac{d \nu}{(2\pi)}
  \frac{v_{q}}{\epsilon(q,i \nu)} \frac{1}{i \nu +\omega-\varepsilon_{\kvec+\qvec}}
\right\} \label{SigmaRPA} \eeq where $\varepsilon_F=k_F^2/2m$
is the unperturbed Fermi energy. From the self-energy, the
renormalization factor, $Z$, at $k=k_F$ is given by \beq
Z^{-1}=1- \frac{ \partial \Sigma(k_F,\omega)}{\partial \omega}
\Big|_{\omega=\varepsilon_{F}} \label{Z} \eeq and the effective
mass, $m_\sigma^*$, which characterizes the quasi-particle
dispersion at the Fermi surface, is then obtained via \beq
\frac{m}{m^*} = Z\left( 1+ \frac{m}{k_F} \frac{\partial
\Sigma(k,\varepsilon_F)}{\partial k} \Big|_{k=k_F} \right) =
\frac{1+ \frac{m}{k_F} \frac{\partial
\Sigma(k,\varepsilon_F)}{\partial k} \Big|_{k=k_F}} {1- \frac{
\partial \Sigma(k_F,\omega)}{\partial \omega}
\Big|_{\omega=\varepsilon_{F}}} \label{mstar} \eeq

Formally, the RPA contains the leading order corrections beyond
the independent particle Hartree-Fock approximation, and its
validity is restricted to the high density region, $r_s \to 0$.
In this limit, it is consistent to expand Eq.~(\ref{mstar}).
This leads to the so-called on-shell approximation (OSA) of the
effective mass, \beq \frac{m}{m^*_{OSA}} = 1+ \frac{m}{k_F}
\frac{\partial \Sigma(k,\varepsilon_F)}{\partial k}
\Big|_{k=k_F} + \frac{ \partial \Sigma(k_F,\omega)}{\partial
\omega} \Big|_{\omega=\varepsilon_{F}}. \label{mstarOSA} \eeq
In the literature, there exists a long-standing controversy
\cite{Rice,lee,Santoro,Sarma,Asgari}, whether Eq.~(\ref{mstar})
or Eq.~(\ref{mstarOSA}) should be used to determine the
effective mass. Strictly speaking, the validity of the RPA is
confined to the high density region where both formulas
essentially agree, $r_s \simle  1$, and the validity of RPA or
RPA-like results at lower densities essentially relies on
cancelation of errors. However, given the exact self-energy
beyond RPA, only Eq.~(\ref{mstar}) provides the effective mass,
and  our  finite-size analysis below is based on
Eq.~(\ref{mstar}) , since, as we will see, our results are not
restricted to the RPA.

\section{Thermodynamic limit extrapolation of finite system results}
All formulas of the previous section are valid for systems of
finite size, $L=V^{1/d}$, where wave vectors are discrete.
Perturbative expressions for the Green's function  and the
dielectric functions, as \emph{e.g.} the RPA expressions, are
based on the linked-cluster theorem, valid in  a
grand-canonical ensemble of fixed volume introducing small
number fluctuations. Since the non-interacting Green's function
does not depend explicitly on the system size, the
extrapolation from the finite to the infinite system
corresponds to the conversions of all underlying discrete sums
to integrals, $(2\pi/L)^d \sum_{\qvec} \dots \to \int d\qvec
\dots$, and finite size corrections are equivalent to
discretization errors in quadrature. On this basis, the
elimination of the size error in the second term on the rhs of
the total energy expression, Eq.~(\ref{Etot}), is
straightforward, leading to the Madelung constant \beq v_M=
\sum_\qvec \frac{v_q}{2V} - \int \frac{d\qvec}{(2\pi)^d}
\frac{v_q}{2}. \eeq In the following, we show that  a similar
analysis of the quadrature error can be used to reduce the
finite size error in other quantities.

If the integrand is regular, the discretization error is
rapidly vanishing  (of order $1/N$ or higher). The main
contribution of order $N^{-\alpha}$ with $\alpha<1$ comes from
non-analytical points in the integrand. From the general
structure of the perturbation expansion there are only certain
discrete points where non-analyticities can develop: in the
long wavelength limit, $k=0$, where the Coulomb potential is
singular, and for integer multiples of the Fermi wave-vector
$k=j\, k_F$ ($j=1,2,\dots$) due to the sharp Fermi surface of
the underlying wave function.

Explicitly, the sharp Fermi surface at $k_F$ leads to
well-known shell oscillations in the energy, already present in
the energy of the ideal Fermi gas, Eq.~(\ref{Ekin}). Twisted
boundary conditions strongly reduce these oscillations
\cite{TABC}, and grand-canonical  twist averaging (GTABC)
completely eliminates them \cite{FSE}. Within GTABC, we  have
\beq E_0^{GTABC}= \frac{2}{N} \int_{|\vartheta_\alpha|\le
\pi/L} \frac{d\vartheta}{(2\pi/L)^d} \sum_{\kvec_\alpha=2\pi m
/L}
\theta(\varepsilon_F-\varepsilon_{\kvec+\vartheta})\varepsilon_{\kvec+\vartheta}
=\frac{2}{n} \int  \frac{d\kvec}{(2\pi)^d}
 \theta(\varepsilon_F-\varepsilon_{\kvec}) \varepsilon_{\kvec}
\eeq and $E_0^{GTABC}\equiv E_0(N\to \infty)$. Similarly, GTABC
also eliminates size effects of other non-interacting
properties, \emph{e.g.} in the ideal gas density response,
$\chi_0^{GTABC}(q,\omega) \equiv \chi_0(q,\omega;N\to \infty)$,
as can be seen from the spectral representation \beq
\chi_0^{GTABC}(q,\omega) = \frac{2}{V}
 \int_{|\theta_\alpha|\le \pi/L} \frac{d\vartheta}{(2\pi/L)^d}
 \sum_{\kvec_\alpha=2\pi m/L} \frac{\theta(\varepsilon_F-\varepsilon_{\kvec+\vartheta})-\theta(\varepsilon_F-\varepsilon_{\kvec+\vartheta+\qvec})}
 {\omega+ \varepsilon_{\kvec+\vartheta} -\varepsilon_{\kvec+\vartheta+\qvec}+i \eta}
 \eeq
although $\chi_0$ is still only given at  discrete wave vectors
($\qvec_\alpha = 2\pi j/L$ with integer $j$). Therefore,
twisted boundary conditions as well as GTABC do not modify the
discrete sum in the exchange-correlation energy on the rhs of
Eq.~(\ref{Etot}). However, GATBC lead to an accelerated
convergence of the integrand towards the thermodynamic limit on
all allowed wave vectors.

Within GTABC, we assume that size effects in the integrands of
all expressions can be neglected, and focus on the presumably
largest contributions from the Coulomb singularity at $k=0$.
For the total energy, Eq.~(\ref{Etot}), we expect that the
leading order size corrections of the exchange-correlation
energy, $\delta E$,  is given by \beq \delta E \simeq
\int_{|\qvec_\alpha|\le \pi/L} \frac{d\qvec}{(2\pi)^d} \int_0^1
d \lambda \int_0^\infty \frac{d \omega}{2\pi n} \left[
\frac{1}{\epsilon(q,i\omega;\lambda)}-1 \right]. \eeq
Similarly, the effect of the thermodynamic limit extrapolation
on the renormalization factor and the effective mass can be
obtained from explicit (approximate) expressions of the
self-energy, Eq.~(\ref{SigmaRPA}), together with Eq.~(\ref{Z})
and Eq.~(\ref{mstar}).

\section{Results}
The Lindhard function, $\chi_0(q,\omega)$, has the following
limiting form  in the long wave length limit \beq
\chi_0(k,\omega) \approx n k^2/m\omega^2, \quad k \to 0 \eeq
and we have \beq \epsilon(k,\omega) \approx 1 -
\omega_p^2(k)/\omega^2, \quad k \to 0 \label{epsilonqsmall}
\eeq where the plasma frequency is given by $\omega_p(k)=(n v_k
k^2/m)^{1/2}$.

\subsection{Total energy}
Inserting the limiting form of the dielectric function in Eq.~(\ref{ERPA}), we obtain
\beq
\delta E=
\frac{1}{n}
\int_{|\qvec_\alpha|\le \pi/L} \frac{d\qvec}{(2\pi)^d}
\int \frac{d \omega}{2\pi} \log (1+ \omega_p^2(q)/\omega^2)
=
\frac{1}{n}
\int_{|\qvec_\alpha|\le \pi/L} \frac{d\qvec}{(2\pi)^d} \frac{\omega_p(q)}{2}
\eeq
and size corrections in the energy have a simple interpretation in terms of
zero point energy of long wavelength plasmon modes. Explicitly, in three dimensions, we have
\beq
\delta E_{3d}=\frac{\hbar^2}{2ma_B^2} \sqrt{\frac{3}{ r_s^3}} \frac{1}{N}
\eeq
in agreement with Ref.~\cite{FSE}, and
\beq
\delta E_{2d}=s_2 \frac{\pi}{5} \left( 4\pi  \right)^{1/4} \frac{1}{r_s^{3/2}} \frac{1}{N^{5/4}}
\eeq
in two dimensions \cite{Momk2D}
where $s_2=\frac5{4\pi}\int_{|q_\alpha|<1} |q|^{1/2}=\frac4\pi\int_0^1\!du\,(1+u^2)^{1/4}\simeq 1.362073657 $ accounts for the cubic geometry.

\subsection{Renormalization factor and effective mass}

For the calculation of the renormalization factor and the
effective mass, we need the derivatives of the self energy at
the Fermi surface. Within the RPA, we have \beq \frac{ \partial
\Sigma(k_F,\varepsilon_F)}{\partial \omega} &=&
-\frac{1}{V}\sum_{\qvec \ne  0} \int_{-\infty}^\infty \frac{d
\nu}{(2\pi)}
 \left[  \frac{1}{\epsilon(q,i \nu)}
 -\frac{1}{\epsilon(q,0)}
 \right]
  \frac{v_q}{[i \nu +\varepsilon_F-\varepsilon_{k_F+\qvec}]^2}
  \\
 \frac{m}{k_F} \frac{ \partial \Sigma(k_F,\varepsilon_F)}{\partial k}
   &=&
  \frac{1}{V}\sum_{\qvec \ne 0} \int_{-\infty}^\infty \frac{d \nu}{(2\pi)}
  \frac{1}{\epsilon(q,i \nu)}
  \frac{v_q}{[i \nu +\varepsilon_F-\varepsilon_{k_F+\qvec}]^2}
  \left[ 1+ \frac{ \kvec_F \cdot \qvec}{k_F^2} \right]
\eeq
Since $\chi_0(q,0)=-dn/\varepsilon_F$ for $q\to 0$,  the static dielectric function diverges
in the long wave length limit,
\beq
\epsilon_{RPA}(q,0) = 1+ d n v_q/\varepsilon_F \sim v_q \sim q^{1-d}, \quad q\to 0
\eeq
providing perfect screening,
and the dominant size correction, $\Delta$, of $\partial \Sigma/\partial \omega$
exactly equals  that of $-\partial \Sigma/\partial k$, and is given by
\beq
\Delta
&=&
-\int_{|\qvec_\alpha| \le \pi/L} \frac{d\qvec}{(2\pi)^d}
 \int_{-\infty}^\infty \frac{d \nu}{(2\pi)}
 \frac{1}{\epsilon(q,i \nu)}
  \frac{v_q}{[i \nu +\varepsilon_F-\varepsilon_{k_F+\qvec}]^2}
  \\
  &\simeq&
  \int_{|\qvec_\alpha| \le \pi/L} \frac{d\qvec}{(2\pi)^d}
 \int_{-\infty}^\infty \frac{d \nu}{(2\pi)}
 \frac{v_q}{1+\omega_p^2(q)/\nu^2}
 \frac{1}{\nu^2}
\eeq
We get
\beq
\Delta
=
  -\int_{|\qvec_\alpha| \le \pi/L} \frac{d\qvec}{(2\pi)^d}
  \frac{v_q}{4 \omega_p(q)}
\eeq
and we have
\beq
\Delta_{3d}=-c_3\left( \frac{r_s}{3} \right)^{1/2} \left( \frac{3}{4\pi N} \right)^{1/3}
\eeq
in three dimension \cite{Momk3D} with $c_3=\frac1{4\pi}\int_{|q_\alpha|<1} |q|^{-2}=\frac6\pi\int_0^1\!du\,\frac{\ln(2+u^2)}{1+u^2}\simeq1.221374804$, and
\beq
\Delta_{2d}=- c_2\left(\frac{r_s}{8} \right)^{1/2} \left(\frac{\pi}{N} \right)^{1/4}
\eeq
in two dimensions \cite{Momk2D} with $c_2=\frac5{4\pi}\int_{|q_\alpha|<1} |q|^{-3/2}=\frac4\pi\int_0^1\!du\,(1+u^2)^{-3/4}\simeq1.057929920$.
The leading order size corrections for the renormalization factor are then given by
\beq
Z_\infty^{-1} \simeq Z_N^{-1} + \Delta
\label{Zdelta}
\eeq
and, using Eq.~(\ref{mstar}), the effective mass correction writes
\beq
\frac{m}{m^*} -1 \Big|_\infty&\simeq&
\left( \frac{m}{m^*} -1 \right)\Big|_N
\left( 1+
 \Delta Z
 \right)
 \label{mdelta}
\eeq Note, that, within the validity of the RPA, we may put $Z
=1$ on the rhs of Eq.~(\ref{mdelta}). The  slow decay of
$\Delta$ with increasing system size can lead to important
qualitative and quantitative changes of spectral quantities
when properly extrapolated to the thermodynamic limit with
respect to calculations which rely on linear extrapolation
schemes.

\section{Discussion}

We have discussed size effects of the total energy, the
renormalization factor, and the effective mass. Particularly,
in two dimensions, $Z$ and $m^*$ show an unexpected slowly
decaying size effects $\sim N^{-1/4}$.
  In general, we expect
that these size effects are also present in other response
quantities, \emph{e.g.} the spin susceptibility, as they are
connected to $m^*$ by relations based on Landau's Fermi liquid
theory. Our results on the finite size corrections are quite
general and do not rely on the validity of the RPA, as they are
based on the large wave length behavior of the dielectric
function, Eq.~(\ref{epsilonqsmall}), which applies to all
metallic systems.
Further, all
results agree with the leading order corrections in
\cite{FSE,Momk2D,Momk3D} where finite size corrections have
been derived from a formulation directly based on the
many-body wave function.

\section*{Acknowledgements}
This research was supported by CNRS-IDRIS, Project IDRIS
061801.

\end{document}